\documentclass[12pt,preprint]{aastex}

\def\arcsec{{$^{\prime\prime}$}}
\def\ptsec{$^{\prime\prime}\mskip-7.6mu.\,$}

\shorttitle{Molecular Hydrogen Orbiting a T Tauri Star}
\shortauthors{Bary et al.}

\begin{document}

\title{Detection of Molecular Hydrogen Orbiting a `Naked' T Tauri Star}

\author{Jeffrey S. Bary and David A. Weintraub} 
\affil{Department of Physics \& Astronomy, Vanderbilt University, P.O. Box 1807 Station B, 
Nashville, TN 37235}
\email{bary@eggneb.phy.vanderbilt.edu and david.weintraub@vanderbilt.edu}
\and
\author{Joel H. Kastner}
\affil{Carlson Center for Imaging Science, RIT, 54 Lomb Memorial Drive, Rochester, NY 
14623}
\email{jhkpci@cis.rit.edu}

\begin{abstract}
Astronomers have established that for a few million years newborn stars
possess disks of orbiting gas and dust. Such disks, which are likely
sites of planet formation, appear to disappear once these stars reach
ages of 5-10 $\times$ 10$^6$ yr; yet, $\ge$ 10$^7$ yr is thought
necessary for giant planet formation.  If disks dissipate in less time
than is needed for giant planet formation, such planets may be rare and
those known around nearby stars would be anomalies.  Herein, we report
the discovery of H$_2$ gas orbiting a weak-lined T Tauri 
star heretofore presumed nearly devoid of circumstellar
material. We estimate that a significant amount of H$_2$ persists in the
gas phase, but only a tiny fraction of this mass emits in the near-infrared.
We propose that this star possesses an evolved disk that has escaped
detection thus far because much of the dust has coagulated into
planetesimals. This discovery suggests that the theory that disks are
largely absent around such stars should be reconsidered.  The
widespread presence of such disks would indicate that planetesimals can
form quickly and giant planet formation can proceed to completion
before the gas in circumstellar disks disperses.  
\end{abstract}


\section{Introduction}

The thick envelopes of dust and gas that surround young stellar objects
evolve into circumstellar disks on a timescale of order 10$^6$ yr 
\citep{stro95a}.  The absence of thermal emission from millimeter or
smaller-sized dust grains, as measured through levels of millimeter and
infrared continuum emission much above that expected from a naked star
\citep{beck96,oste95,andr94}, as well as the non-detection
\citep{zuck95,duve00} of molecular line emission from CO towards
more evolved, weak-lined T Tauri stars (wTTS) studied suggests that the disks
become dust and gas depleted in only 3-5 $\times$ 10$^6$ yr.

Two different explanations may account for these observations.
According to the first, circumstellar material has dissipated through
one or more of several processes: photo-evaporation, accretion onto the star, 
tidal dissipation by nearby stars, outflows, or strong
stellar winds \citep{holl00}.  Alternatively, larger
objects could have formed through accretion, thereby
reducing the total surface area of the dust and burying the molecular
component (in the form of ices) inside growing protocometary grains.
Consequently, the levels of thermal emission from solids and line
emission from molecules would drop below detection thresholds as the
effective surface area of solids and number of molecules in the gas
phase decrease.  Hydrogen, which accounts for $>$70\% of the mass of the
disk, would be collected into planetesimals last, assuming that the
hydrogen can avoid being removed from the disk until the planetesimals
have grown large enough to gravitationally bind gas.

According to the second scenario, hydrogen should remain in the disk
until long after the dust becomes undetectable; however, H$_2$ is
difficult to stimulate due to its lack of a dipole moment and the low
temperatures normally found in circumstellar environments.
Consequently, the bulk of the disk material is nearly impossible to
detect directly, so observers typically resort to emission from dust
and abundant trace molecules for insight into the physical conditions
within and masses of circumstellar disks.  The evidence from these
tracers indicate that classical T Tauri stars (cTTS) are surrounded by
substantial disks of gas and dust, while wTTS ---
similar, but possibly slightly more evolved objects --- are `naked,'
essentially diskless.  Together, these data suggest that T Tauri stars
shed their disks in less than 10$^7$ yr, too short a time for gas giant
planets to form unless the disk is much more massive than a `minimum
mass' nebula \citep{liss93}.

High-resolution spectroscopy of near-infrared ro-vibrational emission
from H$_2$ offers a means to detect hydrogen directly and, hence,
provides a diagnostic of the presence of gas in the near environments
of young stars \citep{wein00}. High velocity (shocked) H$_2$ has been
detected in outflows associated with young stars; however, quiescent
H$_2$ associated with orbiting material in circumstellar disks has been
almost impossible to identify.  Detections of cold H$_2$ gas apparently
about 100 AU from the parent stars were reported recently for three
debris-disk stars \citep{thi01a} and from several Herbig Ae/Be stars
and cTTS stars \citep{thi99,thi01b}, based on low
signal-to-noise observations made at 17 and 28 $\mu$m by the {\it
Infrared Space Observatory (ISO)}.  In much more sensitive observations
made from Mauna Kea at 12 and 17 $\mu$m, Richter et al.\ (2002) were
unable to confirm the detections of H$_2$ reported at 17 $\mu$m by Thi
et al.\ (2001b) for the three stars observed by both groups (GG Tau, AB
Aur, HD 163296).  The Richter et al.\ (2002) results call into question
the other {\it ISO} detections as well.  If the {\it ISO} observations
are correct, the detected gas must be outside of 150 AU in
extended, non-disk components. Thus, the only undisputed observations
reporting the detection of the most abundant gas in circumstellar disks
of young stars are both for the cTTS TW Hya
\citep{wein00,herc02}, for which the presence of a circumstellar disk
was both expected and already established.

In this paper, we report the detection of emission from molecular
hydrogen emission from DoAr 21 (= V2246 Oph), which is classified as a wTTS based
on its H$\alpha$ line equivalent width ($-$0.6 \AA; Bouvier \& Appenzeller
(1992)).  Consistent with its being classified as a wTTS, DoAr 21 remains
undetected to very low sensitivity thresholds in thermal emission from
circumstellar dust \citep{andr94}.  The canonical age for wTTS
is 3-10 Myr old, placing stars like DoAr 21 in the planet building
epoch.  However, Wilking et al.\ (2001) suggest that the young stellar
object population in rho Ophiuchus has a shorter than average disk
survival time and quote an age of 0.4 Myr for the class III sources
(wTTS) they studied.  Whether 0.4 or 4 Myr old, as a wTTS DoAr 21 is
presumed to be beyond the disk accretion phase of pre-main sequence
evolution and shows virtually no signs that any circumstellar material
is present.  The only previous hint of the existence of circumstellar
material around this source is an observed polarization amplitude of
5.8\% in 9\ptsec7 aperture observations at I band \cite{ageo97};
however, optical speckle images presented by Ageorges et al.\ revealed
no extended structure around DoAr~21 at 0\ptsec2 spatial resolution and
to a sensitivity level of K$\leq$12.

\section{Observations}

We obtained high-resolution (R $\simeq$ 60,000) spectra of DoAr 21, centered near
the 2.12 $\mu$m ro-vibrational transition of H$_2$, on 2000 June 20-23,
UT, using the Phoenix spectrometer \citep{hink98} on NOAO's 4 m
telescope atop Kitt Peak, Arizona.  In spectroscopic mode, Phoenix used
a 256 $\times$ 1024 section of a 512 $\times$ 1024 Aladdin InSb detector
array.  Our observations were made using a 30\arcsec\ long,
0\ptsec8 (4 pixel) wide slit, resulting in instrumental spatial and
velocity resolutions of 0\ptsec11 and 5~km~s$^{-1}$, respectively.
Our actual seeing limited spatial resolution was $\sim$ 1\ptsec4.  The
spectra were centered at 2.1218 $\mu$m providing spectral coverage from
2.1167 to 2.1257~$\mu$m.  This spectral region includes three telluric
OH lines, at 2.11766, 2.12325, and 2.12497~$\mu$m that provide an absolute
wavelength calibration.

The DoAr 21 spectrum was obtained in a 3600 s integration. Telluric
calibrations were made from 600 s observations of the A0V star Vega.
Flat-field images were made using a tungsten filament lamp internal to
Phoenix.  Observations were made by nodding the telescope
14\arcsec\ along the slit, producing image pairs that were then
subtracted to remove the sky background and dark current.   The DoAr
21 spectrum was extracted within 0\ptsec7 E-W of the source for a total
beam width of 1\ptsec4, beyond which no emission was detected.  The
spectrum was divided by the continuum and ratioed with that of the
telluric calibrator in order to remove telluric absorption features.

The spectrum presented in Figure 1 has been corrected, though
incompletely, for telluric absorption features and shifted in velocity
to the measured v$_{lsr}$ (3.7 km s$^{-1}$) of the rho Ophiuchi cloud
\citep{kama01}.  Although no direct measurements exist for the velocity
of DoAr 21, all published values of v$_{lsr}$ for sources known to be
in the rho Ophiuchus cloud, at a distance of 160 pc, fall within 1 km
s$^{-1}$ of the velocity of the cloud, with respect to the local
standard of rest. Thus, since DoAr 21 is well established as a member
of this cloud, we have made the assumption that the rest
velocity of DoAr 21 is indistinguishable from the velocity of the
cloud.  This spectrum shows a single, spatially unresolved 
emission line, with a full-width-half-maximum
velocity of 9 km s$^{-1}$, associated with the v = 1$\rightarrow$0 S(1)
transition of H$_2$ at 2.12183 $\mu$m, and centered at the
rest velocity of DoAr 21. We estimate the intensity of this line
emission to be 1.5 $\pm$ 0.15 $\times$ 10$^{-14}$ erg s$^{-1}$ cm$^{-2}$.


\section{Discussion}

Emission from cold H$_2$ is not expected from a supposedly diskless,
wTTS, such as DoAr 21.  Yet, we detected a single
emission line in the spectrum of DoAr 21, the central wavelength of
which matches the stellar rest frame wavelength of the 2.12183 $\mu$m
v=1$\rightarrow$0 S(1) ro-vibrational transition of H$_2$ to within
errors ($\pm$1 km s$^{-1}$).  We conclude that the observed emission is
from gaseous H$_2$ within $\sim$110 AU of DoAr 21.

Could this gas be shock excited H$_2$ in a stellar jet, as often is
seen toward young stellar objects? Such gas is expected to have a
velocity of 30-50 km s$^{-1}$ relative to the star. The fact that the
central line velocity of the H$_2$ falls within $\sim$ 1 km s$^{-1}$ of
the velocity of the CO in the rho Ophiuchus cloud, and implicitly at
the rest velocity of DoAr 21, demands that such a jet would have to be
oriented in a highly unlikely geometry, within $\sim$ 2\degr\ of
perpendicular to our line of sight.  Furthermore, since the detected
emission line is spatially unresolved, the jet would have to be
contained to within 110 AU of the star and be very tightly confined along
the outflow axis, since the conical expansion velocity of an outward rushing
jet is likely greater than 10 km s$^{-1}$. We conclude that, in this case,
shock excitation in a stellar jet is an extremely unlikely stimulating
mechanism for exciting the H$_2$ emission.

Could the H$_2$-emitting gas reside in a tenuous halo, rather than a
disk, surrounding the young star?  DoAr 21 is seen through more than 
6 magnitudes of visual extinction \citep{bouv92}; thus, it is perhaps conceivable
that the H$_2$ emission from DoAr 21 could originate in ambient molecular
cloud gas that is excited by ultraviolet photons and/or X-rays from the star.  The fact
that the H$_2$ line emission is spatially unresolved yet marginally
spectrally resolved is inconsistent with this interpretation, as such an
UV- or X-ray-excited halo should be many hundreds or thousands of AU in
extent and would exhibit a very narrow line profile.  We believe that it is
very unlikely that some kind of excitation of a quiescent interstellar gas cloud
would produce broadened, yet spatially unresolved H$_2$ line emission. 
Furthermore, we have failed to detect H$_2$ emission around other rho
Ophiuchus and Taurus cloud stars that display similarly large values of
A$\rm_V$ and large UV and X-ray fluxes (Bary et al., in prep.)  We therefore conclude
that the H$_2$ emission from DoAr 21 arises in a circumstellar disk and from
within 110 AU of the star.

The cTTS LkCa 15 shows a double-peaked velocity
profile for the H$_2$ line at 2.12183 $\mu$m, with a peak-to-peak
separation of 10~$\pm$~1.5~km~s~$^{-1}$ (Bary et al. 2001; Bary et al.,
in prep).  This double-peaked profile is just that which is expected
for H$_2$ emission orbiting LkCa 15 in a disk seen at an inclination
angle of 34$\degr$$\pm$10$\degr$ (for which the inclination angle has been
determined through interferometric, millimeter wavelength observations
of CO emission; Duvert et al.\ (2000)).  Given the kinematic information from the
emission line and the inclination angle, we determine the bulk of the
emission to be at a distance of 10-30 AU from the star,
in Keplerian motion.  Since the line widths, strengths, and
velocities relative to their stars are all very similar for LkCa 15 and
DoAr 21, the physical mechanism for producing the H$_2$ emission also
is likely very similar in both cases.  We therefore infer that the bulk
of the H$_2$ gas producing the emission observed toward DoAr 21, which
we have already constrained to be within 110 AU of the star, probably
is located at distances of 10-30 AU from the star.  Although the
double-peaked velocity profile is not observed for DoAr 21, the absence
of a detectable double-peaked velocity profile is probably due to the
fact that this disk is viewed at a steeper inclination angle ({\it i}
$>$ 55\degr), such that the double-peaked distribution is below the
velocity resolution of our data and because the gas does not reside
within a few AU of the star, since the absolute velocities would be
quite high at small orbital distances.

What is the mechanism for producing the observed H$_2$ line emission
within this disk?  Ultraviolet fluorescence and X-ray radiation are both
capable of stimulating H$_2$ emission since UV pumping and
X-ray ionization \citep{tine97,malo96,gred95} serve to enhance the
population of the first excited vibrational state of H$_2$ under the right
conditions.  In general, T Tauri stars are good sources of UV
\citep{vale00} and X-ray photons \citep{feig99}.

DoAr 21 is one of the most X-ray luminous T Tauri stars known \citep{casa95}.
DoAr 21 also has significant emission in the UV \citep{bouv92,herb88,safi1995}
above that expected from thermal emission from the photosphere of the star,
with a $U-V$ excess of $-$1.4 magnitudes.  For the last decade, both excess
UV emission and strong X-ray emission \citep{feig99} have been thought of as
indicators of coronal activity for TTS; on the other hand, more recent work suggests
they may be signposts of accretion from a circumstellar disk onto the star
\citep{john00,kast02}.  Kastner et al.\ (2002) model the Chandra X-ray spectrum
of TW Hya and show that the temperature characterizing the bulk of the plasma
producing the X-rays from this star is at a temperature of $\sim$ 3 $\times$
10$^6$ K and that the electron densities are $>$~10$^{12}$~cm$^{-3}$.
This high temperature and high density, they argue, are consistent with
adiabatic shocks produced by gas infalling onto less than 5\% of the surface of
the star at free-fall velocities of a few 100 km s$^{-1}$.  Johns-Krull et
al.\ (2000) compared the level of ``transition region'' line emission from TTS
with main sequence, RS CVn magnetic ``standard'' stars, and wTTS and showed
that cTTS appear to have an anomalous line emission strength with accretion
shocks being the most likely explanation for the excess line strength.  Noting
that DoAr 21 is a strong source of X-rays and UV photons that can be produced
through either accretion processes or coronal activity (with coronal activity
having previously been assumed to be the production mechanism for wTTS because
of their supposed absence of accreting material) and having demonstrated from
our spectra the likelihood that a significant amount of gas still orbits the
wTTS DoAr 21, we suggest that DoAr 21 may be actively accreting material from
its disk.  This suggestion, while contrary to the present paradigm that wTTS
are not accretionally active, is consistent with our detection of UV or X-ray
excited H$_2$ gas in a disk orbiting a wTTS, which is likewise contrary to
present paradigm.  We note, however, that at kT $\sim$ 3 keV, as suggested by
Imanishi et al.\ (2002), the X-ray emission from DoAr 21 may be too hard to
be explained by accretion. 

For X-ray stimulated excitation, the optical depth for X-rays,
determined using cross-sections from Yan et al.\ (1998) and a disk
density profile found in Glassgold et al.\ (2000), is such that the
X-rays must travel through a significant column of gas
(N~$\ge$~10$^{21}$~cm$^{-2}$) before depositing their energy.  Given
the gas densities likely in an inner disk, we find that X-rays produced
in the near-environment of DoAr 21 could penetrate a few tens of AU
into the disk.   Any gas excited by X-ray absorption followed by
collisional excitation in the high density, inner disk will quickly
become collisionally de-excited.  This precludes the X-ray excitation
mechanism from working well in the inner disk; however, this mechanism
may plausibly work at a few tens of AU, especially in lower density
regions at the upper and lower surfaces of the disk, where the density
is below 10$^7$ cm$^{-3}$.  A similar argument holds for UV excited
gas; UV photons cannot penetrate out to tens of AU in the
self-shielding midplane, but they could excite a small volume of gas in
more tenuous regions above the midplane out at 10-30 AU.  UV excitation
would work in the inner disk; however, we should then see a wider H$_2$
line width due to the higher Keplerian velocities of gas within a few
AU of the star.  We cannot definitively determine which
excitation mechanism, X-ray ionization or UV fluorescence, is of primary
importance since both appear to be plausible candidate mechanisms for
producing the observed emission at distances of a few tens of AU from
the star.

For infrared H$_2$ emission, it is straightforward to estimate the mass
of H$_2$ responsible for the detected emission line (see, e.g., Thi et al. 2001a);
however, the mass we calculate (4.4$\times$10$^{-10}$~M$_\odot$) 
will be only a small fraction of the total mass of hydrogen and therefore of the
disk.  Any estimate of the total disk mass depends on a determination of
this fraction.  Our task, then, is to quantify the ratio of the tracer to the total mass.  To
do this, we compared the total disk masses as measured in millimeter CO
line emission and submillimeter thermal emission for the cTTS GG Tau, LkCa 15,
and TW Hya with the masses we estimate from
quiescent, near-infrared line emission of H$_2$.  The amount of H$_2$
gas as determined directly from the line emission at 2.12183 $\mu$m,
and making no assumptions about the gas temperature, for GG Tau and TW
Hya is about 10$^{-7}$ to 10$^{-9}$ of the total mass \citep{bary01}
based on estimates from CO line emission and millimeter and
submillimeter continuum emission.  The amount of H$_2$ gas for LkCa 15
is about 10$^{-8}$ to 10$^{-10}$ of the total mass based on millimeter
and submillimeter continuum emission.  Clearly, mass estimates made
using the line intensity in the v=1$\rightarrow$0 S(1) line greatly
underestimate the true masses in the disks in these three cases, thus
establishing that the warm H$_2$ gas can act as a tracer of much more
massive disks.  Using the ratio of the H$_2$ line-based mass : tracer mass for
GG Tau, TW Hya, and LkCa 15 the H$_2$ line-based mass of 10$^{-10}$
M$_\odot$ for DoAr 21, we can infer a total disk mass range of 10$^{-3}$ to
1 M$_\odot$ for DoAr 21, assuming the 2.12183 $\mu$m H$_2$ line acts as a
tracer of a more massive disk.

Near-infrared and longer wavelength broad band photometry of young
stellar objects previously have been used as methods for identifying
protostellar sources that are likely candidates for harboring
circumstellar disks, as any emission from such stars above that
expected from the nearly isothermal photosphere must emerge from small
dust grains in the near-star environment.  However, DoAr 21 shows no
evidence of excess near-infrared thermal emission
\citep{wilk01}, has at best a marginal mid-infrared excess
\citep{wilk01,bont01}, and has not been detected
in very sensitive 1.3 mm continuum observations \citep{andr94}
(the upper limit for the 1.3 mm flux density of 4 mJy implies the
presence of less than about 10$^{-6}$ M$_\odot$ of dust and
10$^{-4}$ M$_\odot$ of gas). Thus, small dust grains are not present
in sufficient quantities to be detectable and, prior to this work, all
the available evidence suggested that DoAr 21 had no circumstellar disk.

Our detection of a persistent gaseous disk around DoAr 21 is consistent
with recent findings \citep{stas99,rebu01} suggesting that wTTS may not
spin faster than cTTS (but see also Herbst et al. 2000, 2001). Hence,
the discrepancy, if any exists, in X-ray production between cTTS and
wTTS  \citep{feig02} may not be attributable to their spin rates.  The
continued presence of a disk is a logical, though not the only
\citep{hart2002}, explanation for the lack of two distinct
distributions of rotation periods for cTTS and wTTS, since magnetic
decoupling of the star from the disappearing disk, followed by
contraction and spin-up of the star, originally was believed
responsible for the now contested bimodal distribution for T Tauri star
rotational velocities.

The detection of excited H$_2$ from the near-star environment of DoAr 21 is
the first of its kind. Most likely, we have detected only the small annular
portion in the upper layers of a gaseous disk around DoAr 21 that is
sensitive to excitation by X-ray or UV photons.  Thus, the 2.12183 $\mu$m
emission line of H$_2$ likely is tracing the existence of a much larger,
more massive disk, but also one in which significant planet building already
may have occurred.



\clearpage

\begin{figure}
\plotone{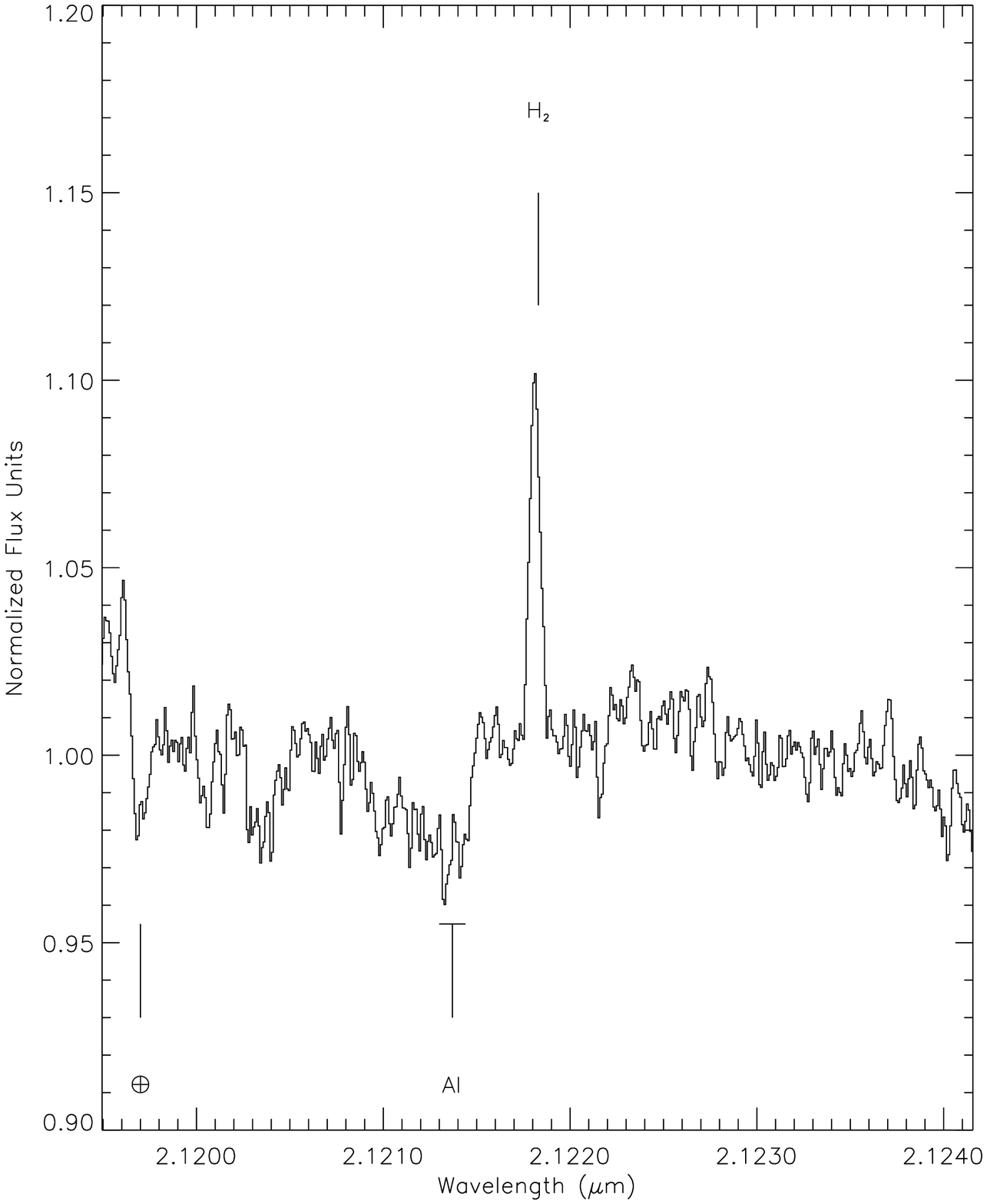}
\caption{Emission from H$_2$ gas, in the v =
1$\rightarrow$0 S(1) transition at 2.12183 $\mu$m, from the disk around
the weak-lined T Tauri star DoAr 21. The absorption feature shortward
of 2.1200 $\mu$m is an incompletely corrected telluric absorption line
while the absorption feature at 2.12137 $\mu$m is tentatively identified
as photospheric 'Al' (Bary et al.\ 2002, in prep) and is labeled as such.
\label{fig1}}
\end{figure}

\end{document}